\documentclass[10pt]{iopart}
\usepackage{iopams} 
\usepackage{graphicx}
\usepackage{epsfig}
\usepackage{float}

\begin{document}
\title[Cosmic expansion history]{Cosmic Chronometers: Constraining the Equation of State of Dark Energy. I: $H(z)$ Measurements}
\author{Daniel Stern$^{1}$, Raul Jimenez$^{2}$, Licia Verde$^{2}$, Marc Kamionkowski$^{3}$ \& S. Adam Stanford$^{4}$\\
\vspace*{.1in}
$^1${\it Jet Propulsion Laboratory, California Institute of Technology, Mail Stop 169-506, Pasadena CA-91109, USA}\\
$^2${\it ICREA \& Institute of Sciences of the Cosmos (ICC), University of Barcelona, Barcelona 08028, Spain}\\
$^3${\it California Institute of Technology, Mail Code 350-17, Pasadena, California 91125, USA} \\
$^4${\it University of California, Davis, CA 95616 and Institute
for Geophysics and Planetary Physics, Lawrence Livermore
National Laboratory, Livermore, CA 94551}\\
\vspace*{.1in}
{\rm E-mail: stern@thisvi.jpl.nasa.gov, raul@icc.ub.edu, licia@icc.ub.edu,
kamion@tapir.caltech.edu, stanford@physics.ucdavis.edu}\\}

\begin{abstract}
We present new determinations of the cosmic expansion history from
red-envelope galaxies. We have obtained for this purpose high-quality
spectra with the Keck-LRIS spectrograph of red-envelope galaxies
in $24$ galaxy clusters in the redshift range $0.2 < z < 1.0$.  We
complement these Keck spectra with high-quality, publicly available
archival spectra from the SPICES and VVDS surveys.  We improve over
our previous expansion history measurements in Simon et al.~(2005)
by providing two new determinations of the expansion history: $H(z)
= 97 \pm 62$~km~sec$^{-1}$~Mpc$^{-1}$ at $z\simeq0.5$ and $H(z) =
90 \pm 40$~km~sec$^{-1}$~Mpc$^{-1}$  at $z \simeq 0.8$. We discuss
the uncertainty in the expansion history determination that arises
from uncertainties in the synthetic stellar-population models.  We
then use these new measurements in concert with cosmic-microwave-background
(CMB) measurements to constrain cosmological parameters, with a
special emphasis on dark-energy parameters and constraints to the
curvature. In particular, we demonstrate the usefulness of direct
$H(z)$ measurements by constraining the dark-energy equation of
state parameterized by $w_0$ and $w_a$ and allowing for arbitrary
curvature. Further, we also constrain, using only CMB and $H(z)$
data, the number of relativistic degrees of freedom to be $4 \pm
0.5$ and their total mass to be $< 0.2$~eV, both at $1\sigma$.

\end{abstract}

\section{Introduction}
\label{sec:intro}

Direct supernova measurements of the deceleration parameter
\cite{Perlmutter:1998np}, as well as indirect
measurements based upon a combination of results from the cosmic
microwave background (CMB)
\cite{cmb}, large-scale structure
(LSS) \cite{percival2df,dodelson}, and the Hubble constant \cite{H0}
indicate that the  expansion is accelerating.  This suggests
either that gravity on the largest scales is described by some
theory other than general relativity and/or that the Universe is filled
with some sort of negative-pressure ``dark energy'' that drives
the accelerated expansion \cite{reviews}; either way, it requires new physics
beyond general relativity and the standard model of particle
physics.  These observations have garnered
considerable theoretical attention as well as
observational and experimental efforts to learn more about the new
physics coming into play.

The simplest possibility is to extend Einstein's equation with a
cosmological constant, or equivalently, to hypothesize a fluid with
an equation-of-state parameter $w\equiv p/\rho = -1$ (with $p$ and
$\rho$ the pressure and energy density, respectively).  However,
it may well be that the cosmological ``constant'' actually evolves
with time, in which case $w\neq -1$, and there are a variety of
theoretical reasons \cite{ratrapeebles} to believe that this is the
case.  Precise measurement of $w(z)$ (with, in general, a parameterized
redshift dependence) or, equivalently, the cosmic expansion history,
has thus become a central goal of physical cosmology \cite{Peacock06,detf}.

Among the techniques to determine the cosmic expansion
history are supernova searches, baryon acoustic oscillations (BAO)
\cite{percival,eisenstein,BAO,pritchard}, weak lensing \cite{weaklensing}, and
galaxy clusters \cite{clusters}.  These techniques all have different strengths,
and they all also suffer from a different set of weaknesses.  As
argued in the ESO/ESA and Dark Energy Task Force reports \cite{Peacock06,detf}, robust
conclusions about the cosmic expansion history will likely
require several avenues to allow for cross checks.  There may
also still be room for other ideas for determining the expansion
history.

A common weakness of supernova searches, BAO (at least the
angular clustering), weak lensing, and cluster-based
measurements is that they rely largely on integrated
quantities.  For example  supernovae probe the luminosity
distance,
\begin{equation}
     d_L(z) =(1+z) \int_z^0 (1+z') \frac{dt}{dz'}dz'.
\label{eqn:dL}
\end{equation}
The other probes rely on similar quantities, which depend on an
integral of the expansion history, to determine the expansion
history, rather than the expansion history itself.  The purpose
of the differential-age technique \cite{jimenezloeb} is to
circumvent this limitation by measuring directly the integrand,
$dt/dz$, or in other words, the change in the age of the
Universe as a function of redshift.  This can be achieved 
by measuring ages of galaxies with respect to a fiducial 
model, thus circumventing the need 
to compute absolute ages. From Galactic globular clusters age-dating we 
know that relative ages are much more accurately determined than
absolute ages (e.g.,
Refs.~\cite{chaboyer,jimenez95,chaboyerkrauss03}).   A
preliminary analysis, with archival data, has already been
carried out \cite{jimenezstern,svj}, and the results applied to
constrain dark-energy theories
(e.g., Refs.~\cite{Hz1}).

The challenge with the differential-age measurement is to find a population of
standard clocks and accurately date them. There is now growing
observational evidence that the most massive galaxies contain
the oldest stellar populations up to redshifts of $z \sim 1-2$
\cite{dunlop96,spinrad,Cowie99,Heavens04,thomas}. Refs.~\cite{Heavens04} and
\cite{panter} have shown that the most massive galaxies have
less than 1\% of their present stellar mass formed at $z <
1$. Ref.~\cite{thomas} shows that star formation in massive systems in
high-density regions --- i.e., galaxy clusters --- ceased by redshift
$z \sim 3$ and Ref.~\cite{treu} shows that massive systems, those with
stellar masses $ > 5 \times 10^{11}$~M$_{\odot}$, have finished
their star-formation activity by $z \sim 2$.
There is thus considerable empirical evidence for a
population of galaxies, harbored in the highest-density regions
of galaxy clusters, that has formed its stellar population at
high redshift, $z > 2$, and that since that thime this population has
been evolving passively, without further episodes of star
formation. These galaxies trace the
``red envelope,'' and are the oldest objects in the Universe
at every redshift.  The differential ages of these
galaxies should thus be a good indicator for the rate of change
of the age of the Universe as a function of redshift.

Here we report the first results of an observational campaign to
obtain high signal-to-noise Keck spectra of red-envelope galaxies
in a number of galaxy clusters at redshift $z<1.0$. In particular,
we report spectroscopy for galaxies in 24 rich galaxy clusters.
These new data extend earlier results from archival field galaxy
data at $z<0.5$ \cite{jimenezstern,svj} to the redshift range
$0.5<z<1$ crucial for determining $w$.  They also allow us to probe
massive galaxies in clusters, which are older, and thus provide a
much more efficient avenue to determine the age at any given redshift.

Our paper is organized as follows.  We first estimate in \S
\ref{sec:statistics} the statistical uncertainties in the age
determination that arise from noisy spectra and finite sample sizes.
The analysis demonstrates that the age and metallicity of galaxies
can be disentangled from the measured spectra given high enough
signal-to-noise, wide enough spectral coverage and accurate modeling,
and it helps identify the wavelength ranges required to optimize
the age determination.  These arguments motivate the Keck observations
we have made.  We discuss the stellar-population models in \S
\ref{sec:stellarpops} and estimate the uncertainties in the age
determination that arise from uncertainties in the stellar-population
models.  Section~\ref{sec:sample} discusses the observations and
the galaxy spectra used in our analysis.  These include both new
spectra obtained at Keck (described in more detail and catalogued
in Paper II, \cite{Stern09b}), as well as galaxy spectra from
archival data.  Section~\ref{sec:ageredshift} presents results for
our measurement of the age-redshift relation, or equivalently, for
the expansion rate $H(z)$ as a function of $z$.  In \S
\ref{sec:cosmoparms}, we combine our measurements of $H(z)$ with
CMB data to derive constraints to cosmological parameters. We
conclude in \S 7.

\begin{figure}
\begin{center}
\includegraphics[width=\columnwidth]{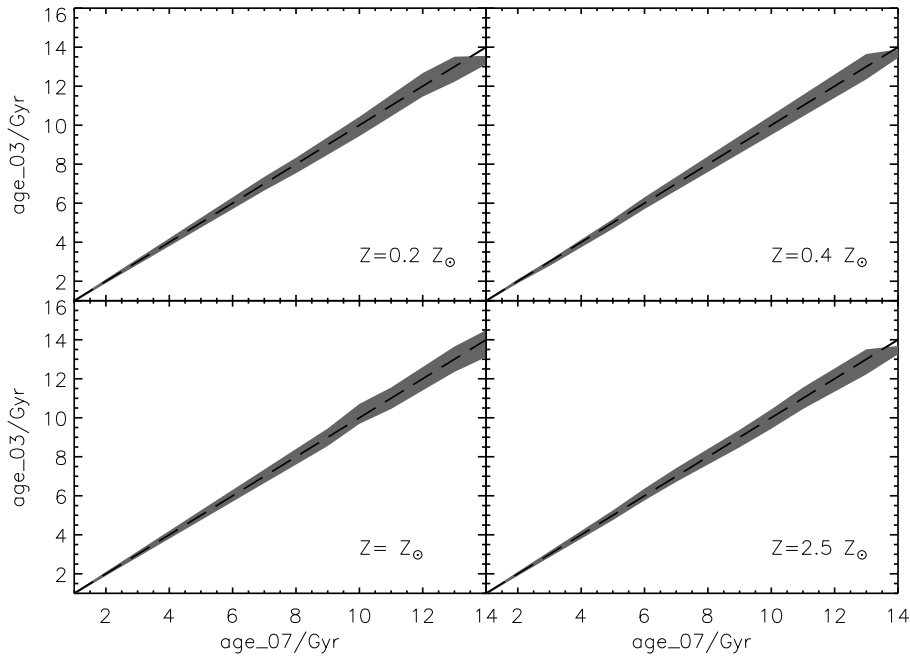}
\caption{\label{fig:fig1}  For sufficiently high $S/N$ data over
     sufficient wavelength range, the age-metallicity degeneracy
     is lifted.  Input mock spectra are from the
     library by CB08 to which we have added random noise with $S/N =
     10$ per resolution element  of 3~\AA. The thin solid line is
     the input age. When we fit the same CB08 spectra to these mock spectra
     with both the age and metallicity as free parameters,
     we find a very good recovery of the age, with a random
     dispersion smaller than a few percent. The age-metallciity
     degeneracy is removed from spectra with blue-light coverage
     ($2500 - 7500$~\AA) and modest $S/N$.}
\end{center}
\end{figure}

\begin{figure}
\begin{center}
\includegraphics[width=\columnwidth]{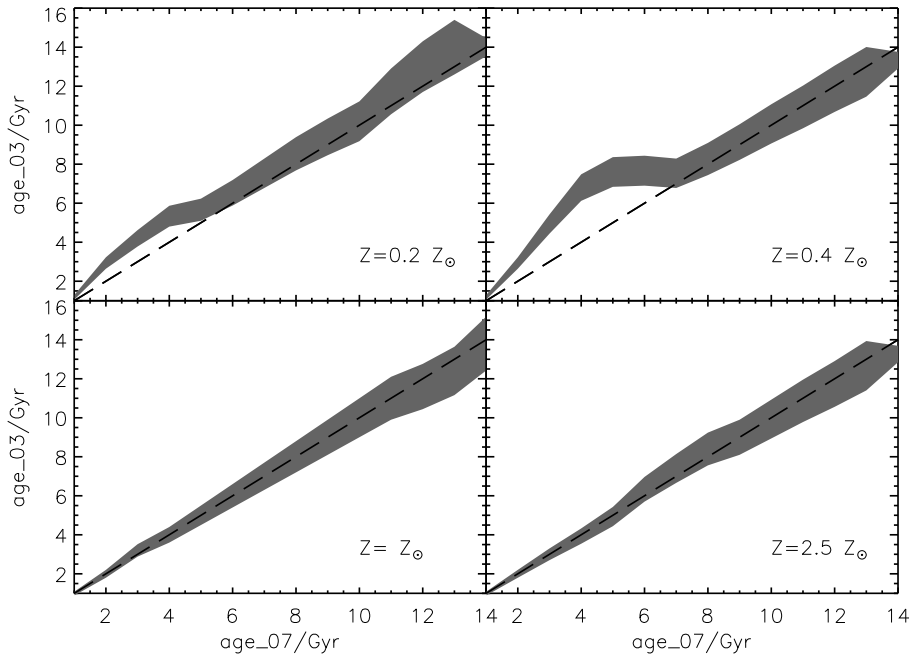}
\caption{\label{fig:fig2} Age recovery for two different
     stellar-population models, CB08 and BC03, which rely on different
     input stellar physics (see text). The (rest frame) wavelength range of
     the fitted spectrum is $2500$ to $7500$~\AA. Note the good
     age recovery despite different input physics, especially at
     higher metallicities where the absorption lines are
     stronger and help lift the age-metallicity degeneracy.} 
\end{center}
\end{figure}

\begin{figure}
\begin{center}
\includegraphics[width=\columnwidth]{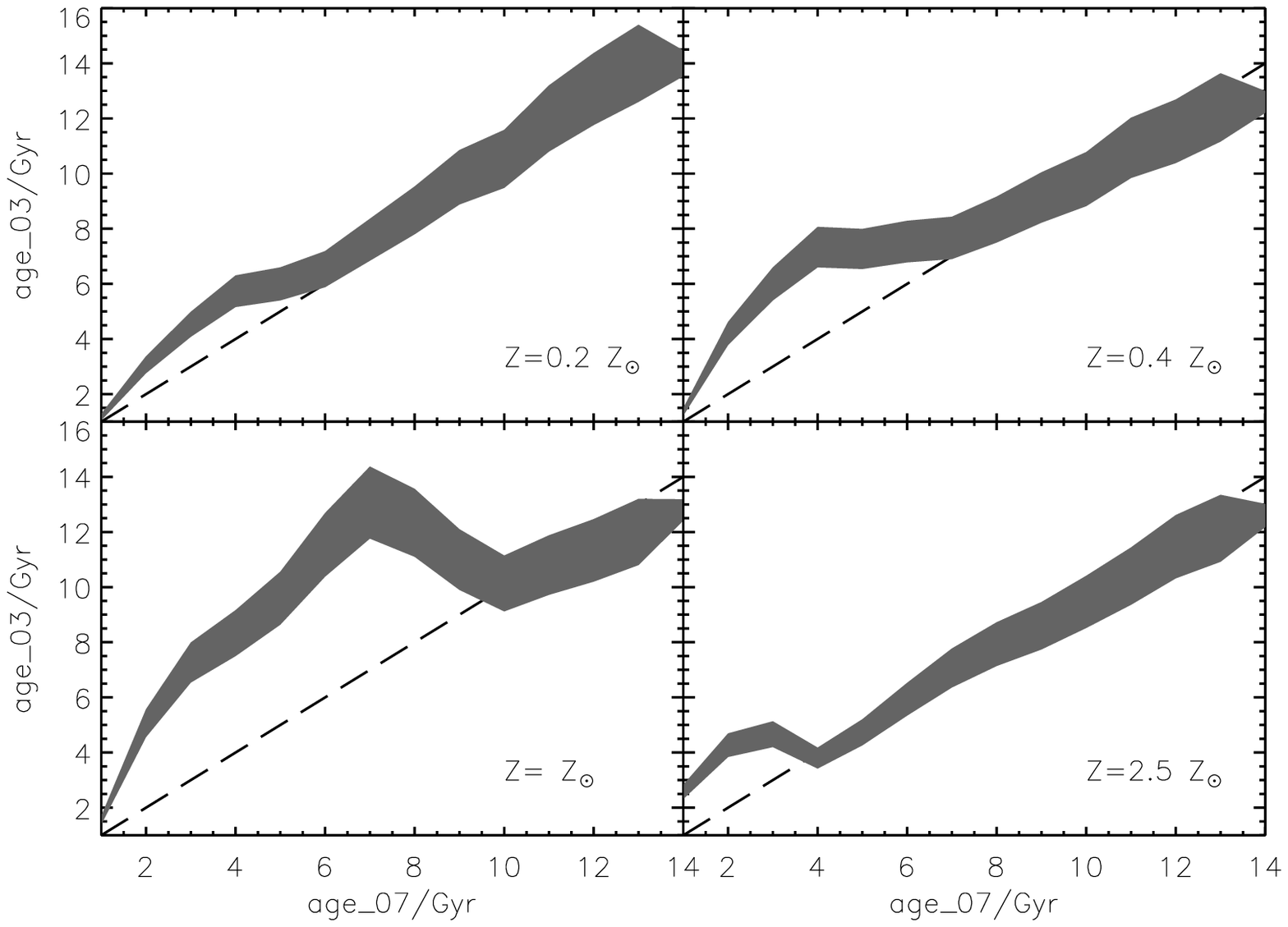}
\caption{\label{fig:fig3} Same as Fig.~\ref{fig:fig2} but for a
     (rest frame) wavelength range of $4500$ to $7500$~\AA. Note that it becomes
     increasingly diffcult to recover the age when blue light in
     the spectrum is not measured.}
\end{center}
\end{figure}

\section{Statistical Uncertainties in the Age Reconstruction}
\label{sec:statistics}

We first quantify the statistical error, arising from noisy
spectra and from finite sample sizes, in the age determination.
To do so we assume that our theoretical stellar models are an
accurate representation of reality and then recover the galaxy
age and metallicity from a spectrum over a given wavelength
range and with a given level of photon noise.  
We have thus generated a series of mock spectra using the  Charlot \& Bruzual (2007) (CB08, Charlot private communication) stellar-population models with
random noise for different values of signal-to-noise and for
different spectral coverages. 
In Fig.~\ref{fig:fig1} we explore how well we recover the age of the
input spectrum for
a signal-to-noise $S/N=10$ per
resolution element of 3~\AA.
The age range is divided into 60 equally spaced bins in age
from $1$ to $14$ Gyr, and 25 equally spaced bins in metallicity from 
from 0.2 to 2.5 $Z_{\odot}$.
In each bin we generate 100 Monte Carlo realizations of the spectrum
with the chosen sigma-to-noise ratio.
For each realization,
parametrized by a given age and metallicity, we maximize the
likelihood in the age and metallicity parameter space and determine the age marginalizing over the metallicity. The metallicity is allowed 
to change from $0.2$ to $2.5$ Z$_{\odot}$. The (rest frame) wavelength range
used goes from $2500$ to  $7500$~\AA.  

By comparing the long-dashed and the solid grey areas, which are
the $1\sigma$ region of the Monte-Carlo simulated  errors, we see
that the recovery is excellent and without bias.  Even for a $S/N$
of only $10$, the age-metallicity degeneracy is lifted when blue
light from (rest frame) $2500$~\AA\ is observed.

\section{Stellar-Population Uncertainties}
\label{sec:stellarpops}

We next investigate the uncertainty that arises in the age
determination from uncertainties in the stellar-population
models.  To do so, we take the mock spectra from \S
\ref{sec:statistics} with the CB08 models and then  recover their age 
(marginalizing over the metallicity) with a different set of
stellar-population models.
Here, we use the Bruzual \& Charlot 2003 \cite{BC03} models which rely on a different set of stellar interior models (Geneva instead of Padova used by CB08) and different stellar atmosphere models. It should be noted that this exercise will generally
{\em over}-estimate the uncertainty, because the models may have
treatments of different phases of stellar evolution that are not
up to date, or because some may be missing phases of stellar
evolution \cite{charlot96}.  As models get updated, they will
hopefully converge toward a more unified picture \cite{Maraston08}.  

Figs.~\ref{fig:fig2}--\ref{fig:fig3} show the comparison
between the two different models for different metallicities
and for different wavelength coverages.   As in Fig. \ref{fig:fig1}, the shaded grey regions correspond to the $1\sigma$ region from $100$ Monte-Carlo realizations. Fig.~\ref{fig:fig2} uses
a (rest frame) spectral range from $2500$ to $7500$~\AA\, while Fig.~\ref{fig:fig3} drops the blue light 
and covers  only the range from (rest frame) $4500$ to $7500$~\AA. In both
cases the $S/N$ is $20$ per resolution element of
3~\AA. Inspection of Fig.~\ref{fig:fig2} shows that there is
good agreement in the age recovery for all ages and
metallicities despite the different stellar input physics in the
models. The recovery is especially good at higher metallicities,
which corresponds to the typical metallicity of the old, massive galaxies
found in clusters. This is expected because the stronger
metallic absorption lines provide a better lever-arm to
determine the metallicity and thus break the age-metallicity
relation. Note that the new CB08 models like Maraston
\cite{maraston} and SPEED \cite{jimenez,Jimenez+98}, apply a
correction to properly model the giant branches.

\begin{figure}
\begin{center}
\includegraphics[width=\columnwidth]{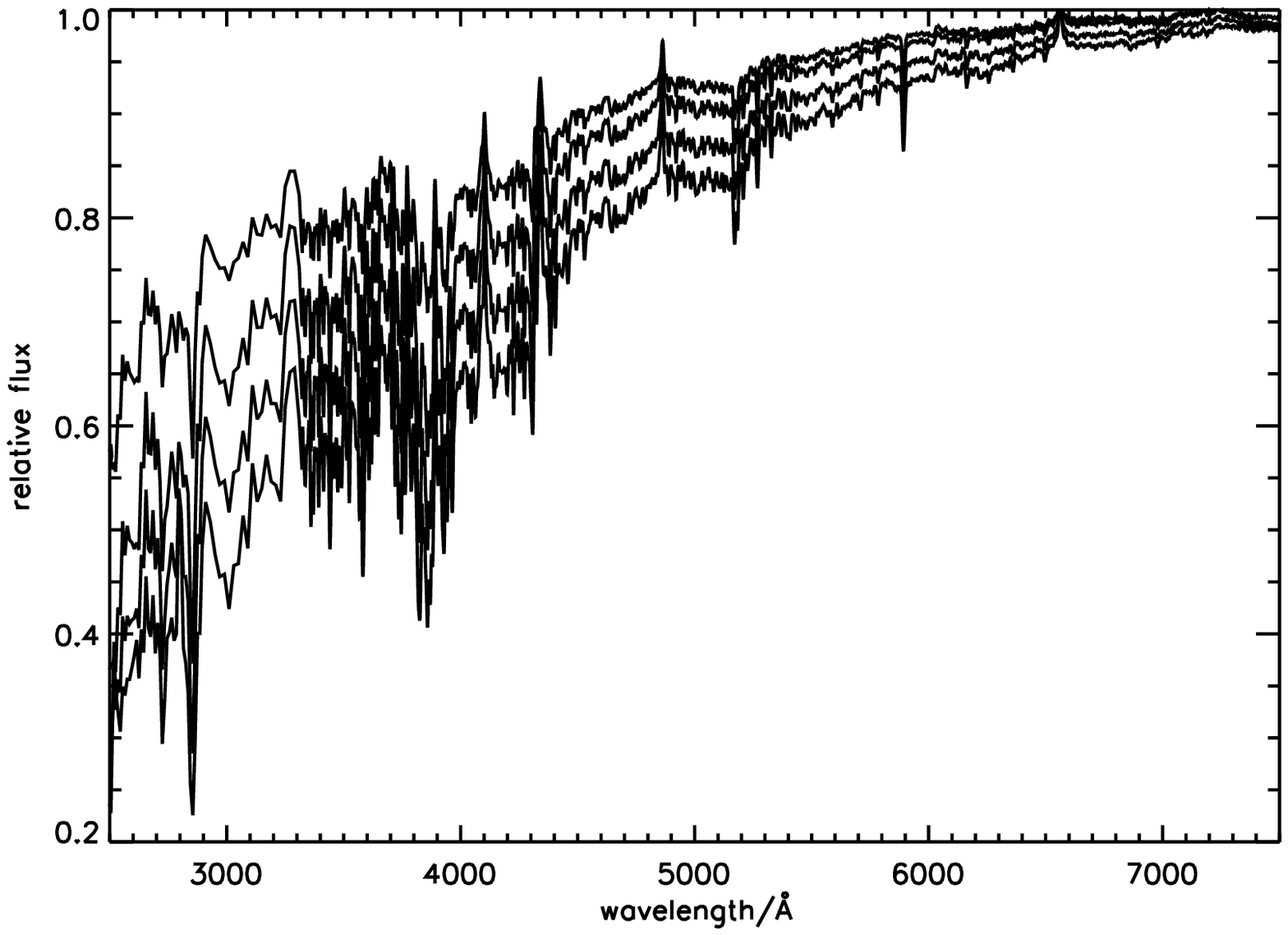}
\caption{\label{fig:figsyn} Relative flux of theoretical
     single stellar populations (SSPs) in
     the wavelength range $2500$ to $7500$~\AA\, for four
     different ages: $3, 5, 7, 9$ Gyr, from top to bottom as a
     ratio of the 2 Gyr spectrum. The plot illustrates how
     differential ages are measured and the essential role of
     blue wavelength coverage for this experiment.}
\end{center}
\end{figure}

\subsection{Importance of blue wavelength coverage}

It is  apparent from comparison of Figs. \ref{fig:fig2} and \ref{fig:fig3} that the wavelength coverage of the
observations is crucial, as the agreement degrades quickly with
observations that do not include light bluewards of the
$4000$~\AA\, break.
The blue end is particularly
important, as line-blanketing at the blue end allows a more
effective breaking of the degeneracy between the age and
metallicity.  It is this consideration that motivates the use of the
Low Resolution Imaging Spectrometer (LRIS, \cite{Oke95}) on the Keck~I telescope, which
is senstive down 
to $\sim3200$~\AA. 

Because the method relies only on measuring relative ages, it is
useful to illustrate the sensitivity of the observed spectrum to
this parameter. In Fig.~\ref{fig:figsyn} we show the relative
flux of spectra for different ages with respect to a 2 Gyr old 
solar metallicity population. All spectra have been arbitrarily
normalised to unity at $7500$~\AA. Even by eye the differences
at a single wavelength are notable, above 10\%. The figure also
illustrates how the greatest sensitivity comes from observations at the blue end of
the spectra, $\approx 3000$ \AA.

\begin{table}
\begin{center}
\caption{Galaxy clusters observed.}
\begin{tabular}{lcccl}
\hline
Galaxy Cluster   &    RA (J2000)  &   Dec (J2000)   &  $z$  & UT Date \\ 
\hline
MS 0906.5+1110      & 09:09:12.7 &   +10:58:29 & 0.172 &  2005 Feb 10    \\
MS 1253.9+0456      & 12:56:00.0 &   +04:40:00 & 0.230 &  2008 Jul 1     \\
Abell 1525         & 12:21:57.8 & $-$01:08:03 & 0.260 &  2007 Dec 18     \\
MS 1008.1$-$1224    & 10:10:32.3 & $-$12:39:52 & 0.301 &  2007 Dec 17     \\
CL 2244$-$0205      & 22:47:13.1 & $-$02:05:39 & 0.330 &  2007 Dec 17-18  \\
Abell 370          & 02:39:53.8 & $-$01:34:24 & 0.374 &  2007 Dec 18     \\
MACS J1720.2+3536   & 17:20:12.0 &   +35:36:00 & 0.389 &  2008 Sep 3      \\
CL 0024+16          & 00:26:35.7 &   +17:09:45 & 0.394 &  2007 Dec 18     \\
MACS J0429.6$-$0253 & 04:29:41.1 & $-$02:53:33 & 0.400 &  2008 Sep 3      \\
MACS J0159.8$-$0849 & 01:59:00.0 & $-$08:49:00 & 0.405 &  2008 Sep 3      \\
Abell 851          & 09:43:02.7 &   +46:58:37 & 0.405 &  2007 Dec 17     \\
GHO 0303+1706       & 03:06:19.1 &   +17:18:49 & 0.423 &  2005 Feb 10     \\
MS 1621.5+2640      & 16:23:00.0 &   +26:33:00 & 0.428 &  2008 Jul 1      \\
MACS J1610.6+3810   & 16:21:24.8 &   +38:10:09 & 0.465 &  2008 Sep 3      \\
MACS J0257.1$-$2325 & 02:57:09.1 & $-$23:26:06 & 0.505 &  2008 Sep 3      \\
MS 0451.6$-$0306    & 04:54:10.8 & $-$03:00:57 & 0.539 &  2005 Feb 10     \\
MS 0451.6$-$0306    & 04:54:10.8 & $-$03:00:57 & 0.539 &  2007 Dec 17-18  \\
Bo\"otes 10.1      & 14:32:06.0 &   +34:16:47 & 0.544 &  2008 Sep 3      \\
CL 0016+16          & 00:18:33.5 &   +16:25:15 & 0.545 &  2007 Dec 18     \\
MACS J2129.4$-$0741 & 21:26:46.9 & $-$07:54:36 & 0.570 &  2008 Jul 1      \\
MACS J0025.4$-$1222 & 00:25:09.4 & $-$12:22:37 & 0.578 &  2008 Jul 1      \\
MACS J0647.7+7015   & 06:47:50.5 &   +70:14:55 & 0.591 &  2009 Mar 2      \\
MACS J0647.7+7015   & 06:47:50.5 &   +70:14:55 & 0.591 &  2009 Mar 3      \\
MACS J0744.8+3927   & 07:44:52.5 &   +39:27:27 & 0.697 &  2009 Mar 2      \\
MACS J0744.8+3927   & 07:44:52.5 &   +39:27:27 & 0.697 &  2009 Mar 3      \\
RCS 2318+0034       & 23:18:31.5 &   +00:34:18 & 0.756 &  2008 Sep 3      \\
MS 1054.5$-$0321    & 10:56:59.5 & $-$03:37:28 & 0.828 &  2002 Mar 11     \\
Bo\"otes 10.8      & 14:32:06.0 &   +34:16:47 & 0.921 &  2008 Jul 1      \\
\hline
\end{tabular}
\end{center}
\label{table:clusters}
\end{table}

\section{Galaxy Spectra Sample}
\label{sec:sample}

\subsection{Keck observations}

\begin{figure}
\begin{center}
\includegraphics[width=0.8\columnwidth]{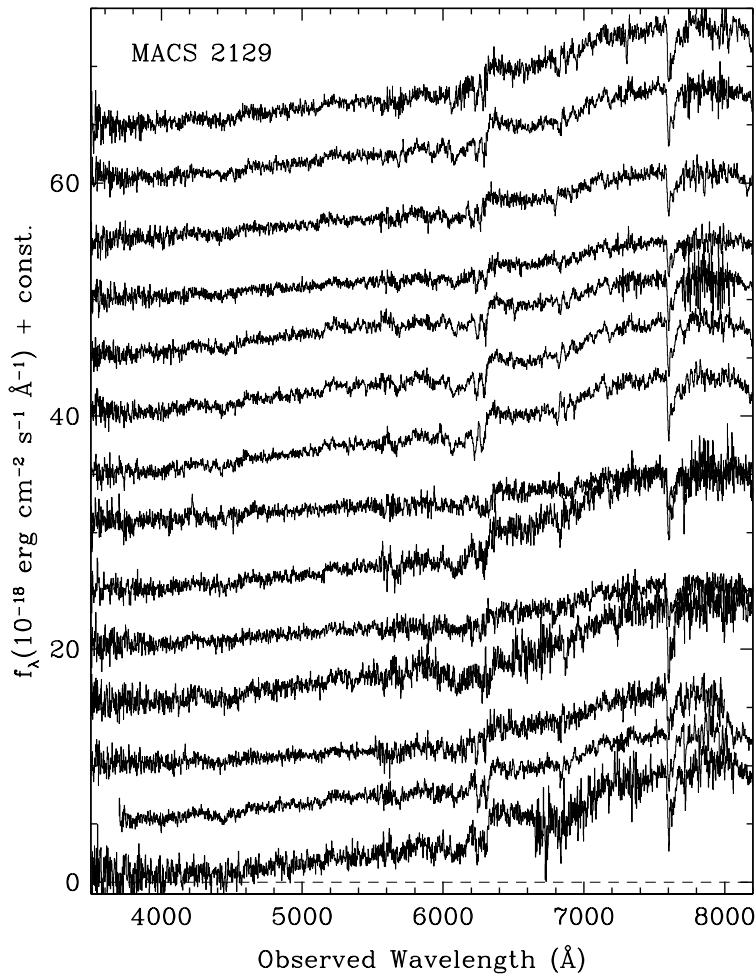}
\caption{\label{fig:specobs} Example of spectra from our observing
program. Observed Keck spectra of red galaxies in the cluster
MACS~J2129.4$-$0741 at $z=0.570$.}
\end{center}
\end{figure}

\begin{figure}
\begin{center}
\includegraphics[width=\columnwidth]{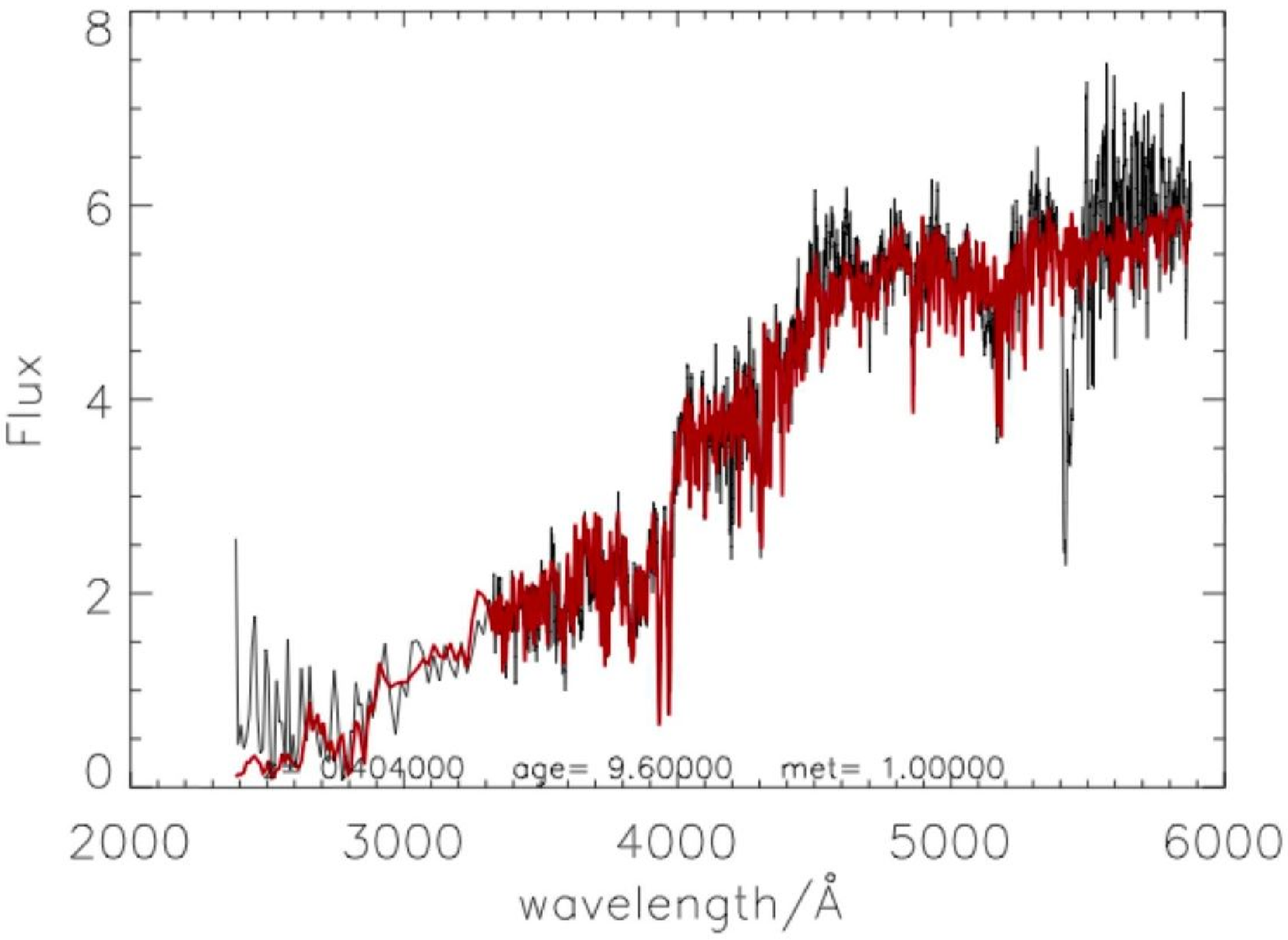}
\caption{\label{fig:specfit} Example of a the model fit (red line)
to observed spectra (black line) on the red envelope for a galaxy
in the cluster MACS~J0429.6$-$0253 at $z = 0.400$.  The best fit
model has solar metallicity and an age of 9.6~Gyr.}
\end{center}
\end{figure}

Motivated by the above considerations, we have started a program
to obtain spectra of bright cluster galaxies with the Keck/LRIS
instrument.  We targeted rich galaxy clusters in order to obtain a
sample as large as possible of red galaxies at redshift $z < 1$.
Most of the clusters are well-known, rich X-ray clusters from  a
variety of samples such as the Abell catalog \cite{Abell}, the ROSAT
Cluster Survey (RCS; \cite{Rosati}) and the Massive Cluster Survey
(MACS; \cite{Ebeling}). In the redshift range $0.5 < z < 1$, fewer
rich X-ray clusters are known, so we targeted (and confirmed) two
of the richest {\it Spitzer} mid-infrared selected cluster candidates
from the IRAC Shallow Cluster Survey \cite{Eisenhardt}.  Many of
the targeted clusters were also observed in the near-infrared cluster
survey of \cite{Stanford}, which provided a valuable and consistent
astrometric resource for slitmask designs. The list of clusters
targeted is presented in Table~\ref{table:clusters} and an example
of observed spectra is shown in Fig.~\ref{fig:specobs}.

The key goal of this program is to provide high--signal-to-noise,
wide-wavelength-coverage spectroscopy of a large number of
early-type galaxies at moderate redshifts. These spectra are
then modeled to derive the ages of the galaxy stellar
populations. Since rest-frame UV light probes the youngest and most
massive stars in a galaxy, blue sensitivity is crucial for this
experiment. Of all the optical spectrographs on $8 - 10$m class
telescopes currently available, LRIS on the Keck I telescope is
unique in being the only dual-beam spectrograph, thus providing
sensitive observations across the entire optical window
($\lambda  \sim 3200$~\AA $- 1 \mu$m).

We obtained the blue channel data with the $400$-line mm$^{-1}$
grism, which has a central wave-length of $3400$~\AA\, and a
spectral range of $4450$~\AA. We obtained the red channel data
with the $400$-line mm$^{-1}$ grating, which has a central
wavelength of $8500$~\AA\, and a spectral range of $3800$~\AA. 
Combining the blue- and red-channel data from the two dichroic
settings, sources typically had final
spectra which spanned the entire $\sim 3200$~\AA$-1 
\mu$m optical window, albeit with higher noise  at the short and
long wavelength extremes. Based on analysis of sky lines,
sources filling the 1.5'' 
wide slitlets used for these observations have resolution $\lambda / \Delta \lambda$ 
$\sim 500$ and $\sim 650$ for the blue and red channels, respectively. Standard stars from \cite{MS90} were observed with the 
same instrument conÞguration for the purposes of spectrophotometric calibration. 
Observations were generally obtained with two dithered exposures
per dichroic conÞguration with typical integration times of
$900$ s to $1800$ s, depending on the cluster redshift and
observing conditions. This allowed both improved cosmic ray
rejection and, by pair-wise subtraction of the images, removal
of the fringing which strongly affects the long wavelength
($\lambda > 7200$~\AA) LRIS data. This required minimum slitlet
lengths of approximately 10''. Since LRIS has an atmospheric
dispersion corrector, mask position angles were optimized based
on the cluster orientation and no special attention was
necessary to align the masks with the parallactic angle.  We
show in Fig.~\ref{fig:specobs} an example of our observed
spectra. See the companion paper \cite{Stern09b} for more details
of the observations, reductions, as well as a catalog of all
$\approx 500$ redshifts measured by this project.

\subsection{SDSS, SPICES and VVDS spectra}

We have extended our Keck observations with the following datasets.
We take advantage of the improvements in calibration provided by
the SDSS spectroscopic pipeline over our previous study
\cite{jimenezstern}, which was based on the early data release (EDR)
sample, and we take advantage of the increased number of luminous
red galaxies (LRGs) available in the public Data Release 6 (DR6)
due to the larger survey area. In addition, we supplement our analysis
using publicly available, high quality data from several surveys that
targeted LRGs at higher redshift than the SDSS.
The SPICES sample \cite{spices}
targeted infrared-selected galaxies ($K < 20$) in about 1/3 square
degree of the sky. Most of the spectra were taken at the Keck
telescope. The VVDS \cite{vvds} is the VIMOS-VLT survey carried out
by the VLT/ESO telescope that spectroscopically targeted galaxies
in the range $17.5 \le I_{\rm AB} \le 24$. Neither of these surveys
was aimed purely at LRGs, and so their efficiency at providing
passively-evolving galaxies is significantly smaller than our
Keck-survey project presented here.

\begin{figure} \begin{center}
\includegraphics[width=\columnwidth,height=20cm]{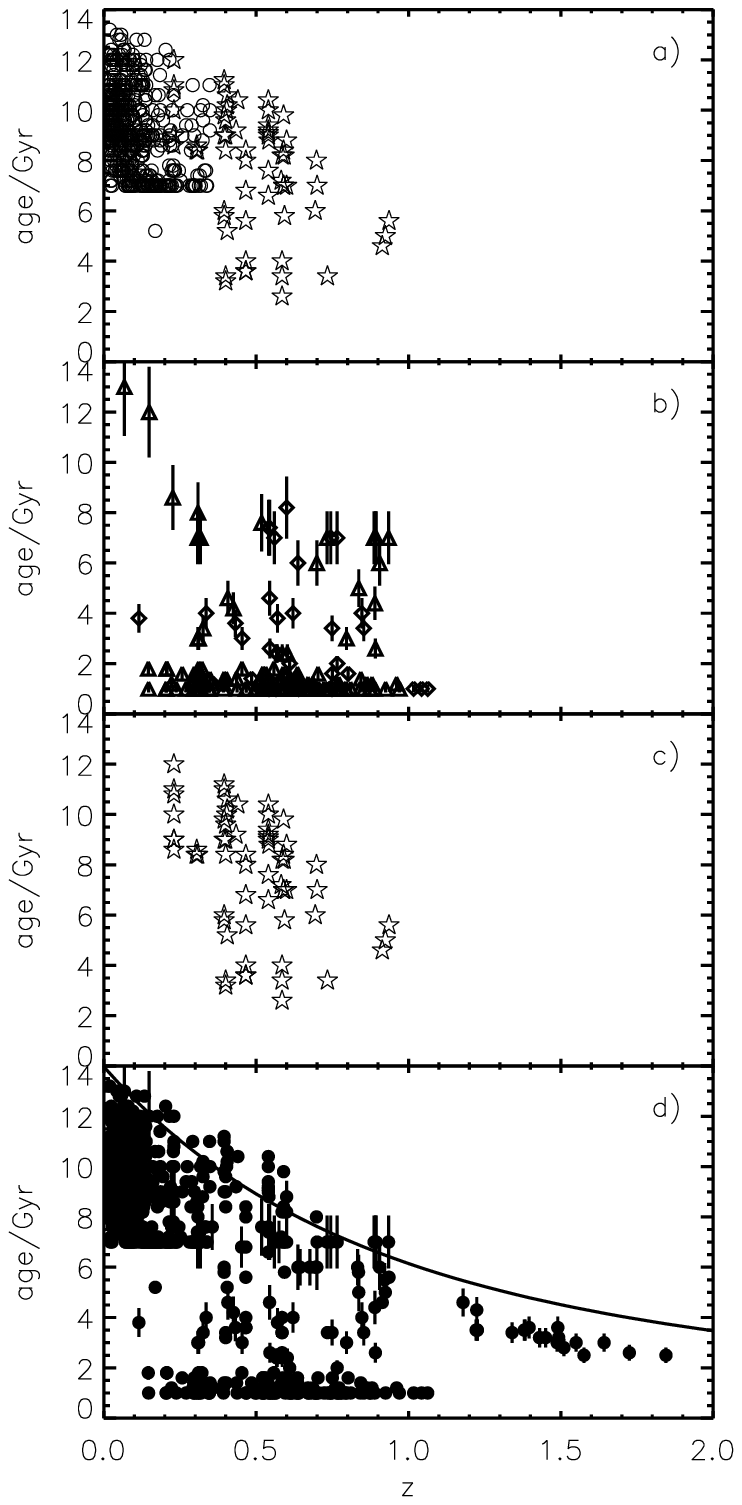}
\caption{\label{fig:agez} Age-redshift relation for (a) SDSS-LRG
     galaxies and Keck clusters; (b) SPICES (diamonds) and VVDS
     (triangles) galaxies; (c) Keck clusters; and (d) all samples,
     including the ones in Ref.~\cite{svj}. The determination of
     $H(z)$ depends only on the determination of the upper (red)
     envelope; younger galaxies do not play any role. Note that
     there is a clear age-redshift relation for  all samples. The
     solid line is the theoretical age redshift relation for the
     $\Lambda$CDM model.}
\end{center} \end{figure}

\begin{figure}
\begin{center}
\includegraphics[width=.8\columnwidth,height=7.5cm]{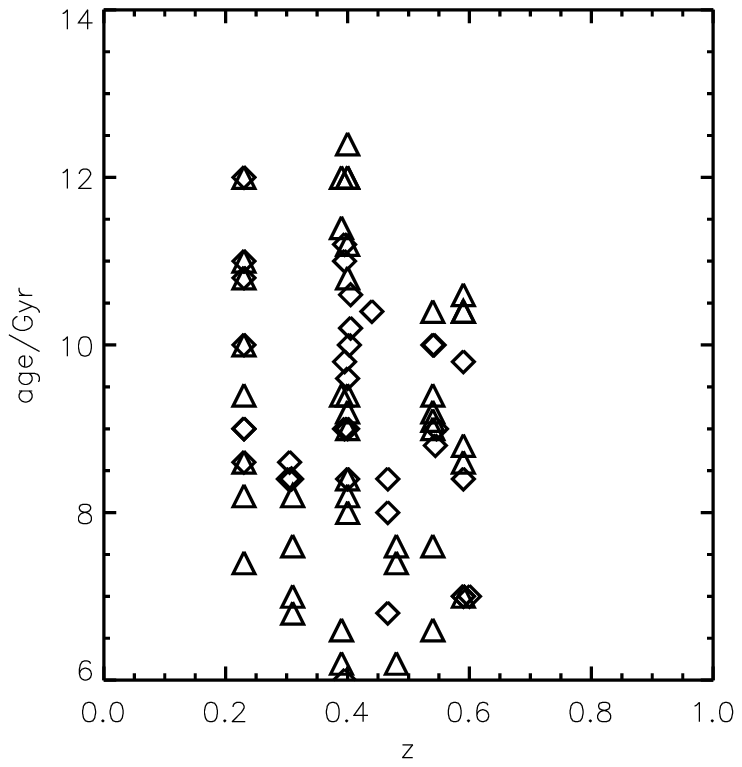}
\caption{\label{fig:comp_clust} Ages obtained with the BC03
     models (diamonds) and the CB08 models (triangles). The plot
     shows that the red-envelope in the age-redshift relation is
     not compromised by the choice of stellar population
     model. The solid line is the theoretical age redshift
     relation for the $\Lambda$CDM model.}
\end{center}
\end{figure}

\section{Measurement of the Age-Redshift Relation}
\label{sec:ageredshift}

We now proceed to measure $dt/dz$ as a function of redshift.
We fit single stellar-population spectra from the
BC03 library  for the full range of age and
metallicity. We choose to use the BC03 library because it is
the most widely used and eases the task of other researchers to
reproduce our results. Fig.~\ref{fig:specfit} shows an example of a model fit to an 
observed spectrum of a galaxy on the red envelope in the cluster MACS~J0429.6$-$0253.  However, significant improvements in
modeling stellar populations have taken placed between the
release of BC03 and now, with new stellar libraries, new models
and new differential methods that have improved the fit of the
models to the data. This is an active field where researchers
are converging to a ``standard'' model;  until such a model is
uncovered, the standard remains BC03. However, as we show
below, the choice of stellar population model does not change
the differential age measurements for the red envelope. We generate single stellar population (SSP)
models with BC03 for ages of $0.1 - 15$ Gyr and
metallicities of $0.1 - 2.5\, Z_{\odot}$. To obtain the best age, we
perform a $\chi^2$ fit
and marginalize over the metallicity to obtain error bars.
Star formation is assumed to occur in a single
burst of duration much shorter ($< 5\%$) than the current age of
the galaxy. We discard fits that have a reduced-$\chi^2$ value
$> 1.2$, as experience has shown that these galaxies have data issues (e.g., poor
$S/N$ ratio due to a bright moon and/or poor weather, poor sky subtraction, contamination from nearby --- and, in 
four cases, 
lensed --- sources,
contamination from an active nucleus), 
or are better fit by an extended star-formation
history and therefore do not belong to the category of passively
evolving stellar populations. We also allow for dust in the fit
following Ref.~\cite{panter} and discard 
those galaxies that contain dust in their fits; i.e., those
where the dust parameter used in Ref.~\cite{panter} is
nonzero. To compute the differential ages, we use the solar metallicity, 
5 Gyr-old model as our reference and
compute $dt$ with respect to it.

The results of the age determination are presented in
Fig.~\ref{fig:agez}, where the four panels show the age-redshift
relation for SDSS-LRG galaxies and Keck clusters (upper panel),
SPICES (diamonds) and VVDS (triangles) galaxies (middle-upper
panel), just Keck clusters (middle-lower panel), 
and all samples including the ones in Ref.~\cite{svj} (bottom
panel). Typical errors on the age are shown for in the
middle-upper panel to avoid cluttering the plots. We also
show in Fig.~\ref{fig:comp_clust} a comparison of the
age-redshift relation for the Keck cluster sample using BC03
(triangles) and CB08 (diamonds) models. Besides two discrepant
galaxies, which in any case are consistent within the errors
with the red envelope edge, the defined edge is nearly
identical. This illustrates that the edge does not overly depend on the
synthetic stellar population model.
 
Note that there is a clear age-redshift relation 
for all samples, either taken individually or together. For
comparison, the solid  line shows  the theoretical age redshift
relation for the standard  $\Lambda$CDM model.  Also note the
excellent performance of the clusters observed with Keck at
selecting the oldest stellar populations, effectively from just
two nights of data in good conditions.  The ages derived from
the Keck clusters nicely fill in the gap in the redshift range
$0.3 < z< 1.0$. Finally, we note that the  different data
samples  are consistent with each other.

We are interested in computing the ``edge'' of this distribution
in the age-redshift plane. As demonstrated in
Ref.~\cite{Jimenezetal03},
this is most efficiently done by computing where the
distribution, as fitted by a one-sided Gaussian, drops by 50\%.
The new data are all at $z <1$.
While the recovered
age-redshift relation is in good agreement with the SDSS one at
$z < 0.3$, the few  additional data points do not change the
determination used in Ref.~\cite{svj}.  Thus, we concentrate
here on the redshift
range  $0.35 < z < 1.0$. We sub-divide  it  in two intervals:
$0.35 < z < 0.6$ and $ 0.6 < z < 1.0$. The second interval is
wider because it is populated by fewer galaxies.  Each of these
intervals is divided in two equal  bins where the edge of the
age-redshift relation is found in a way similar to that
described in Ref.~\cite{Jimenezetal03}, and we also determine
$H(z)$ following Ref.~\cite{Jimenezetal03}. We have checked that the
$H(z)$ determination is robust, within the error bars, to the
choice of binning and the choice of galaxies that populate the edge.

\begin{figure}
\begin{center}
\includegraphics[width=0.8\columnwidth]{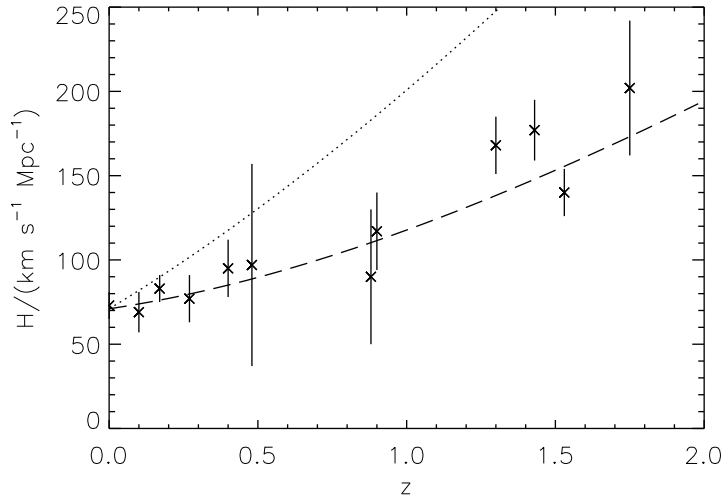}
\caption{\label{fig:Hz}The $H(z)$ measurement from our sample
     of passively evolving galaxies. The points at $z > 1$ are
     taken from Ref.~\cite{svj}, while the point at $z=0$ is
     from the Hubble Key Project \cite{H0}. The dashed line is the $\Lambda$CDM
     model while the dotted line is an Einstein-de-Sitter model.}
\end{center}
\end{figure}

\begin{table}
\begin{center}
\caption{$H(z)$ from passively evolving galaxies.}
\vspace*{0.5cm}
\begin{tabular}{cc}
\hline
& $H(z)$ \\
$z$ & [km sec$^{-1}$ Mpc$^{-1}$] \\
\hline
0   & $73 \pm 8$ \\
 0.1 & $69 \pm 12$   \\
 0.17 & $83 \pm 8$\\
  0.27 & $77 \pm 14$ \\
   0.4 & $95 \pm 17$ \\
    0.48 & $97 \pm 60$ \\   
     0.88 & $90 \pm 40$ \\
     0.9 & $117 \pm 23 $ \\
      1.3 &  $168 \pm 17$ \\
       1.43   & $177 \pm 18$  \\
        1.53  & $140 \pm 14$  \\
         1.75   & $202 \pm 40$  \\
\hline
\end{tabular}
\end{center}
\end{table}

\section{Constraints on Cosmological Parameters}
\label{sec:cosmoparms}

We now proceed to place constraints on cosmological parameters using
the  above determination of $H(z)$ in combination with the
latest results from the WMAP 5-year release \cite{WMAP5}.  For some
applications we also consider the latest determination of
$H_0$ from  Ref.~\cite{Riess09}.   The $H(z)$ data help constrain
cosmological parameters beyond the standard $\Lambda$CDM model,
in particular for parameters that affect the Universe expansion
history or that have degeneracies with expansion-history
parameters. We find improved constraints for
models with arbitrary curvature and  dark-energy
equation-of-state parameter $w \ne -1$, and for models with
non-standard neutrino properties.

\begin{figure}
\begin{center}
\includegraphics[width=0.8\columnwidth]{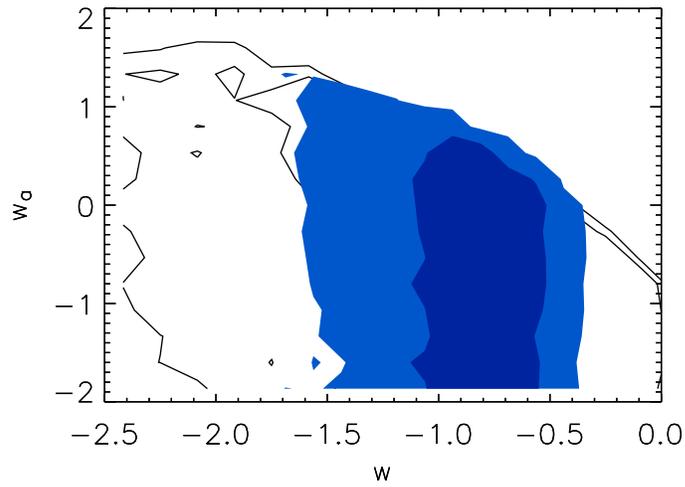}
\caption{\label{fig:wwa} Constraints in the $w$-$w_a$ plane for
     an open-w(z)CDM where $w(z)=w+w_a(1-a)$. The solid empty
     contours are for a model using WMAP5 and $H_0$
     constraints. The filled contours use $H(z)$ information.} 
\end{center}
\end{figure}

\begin{figure}
\begin{center}
\includegraphics[width=0.8\columnwidth]{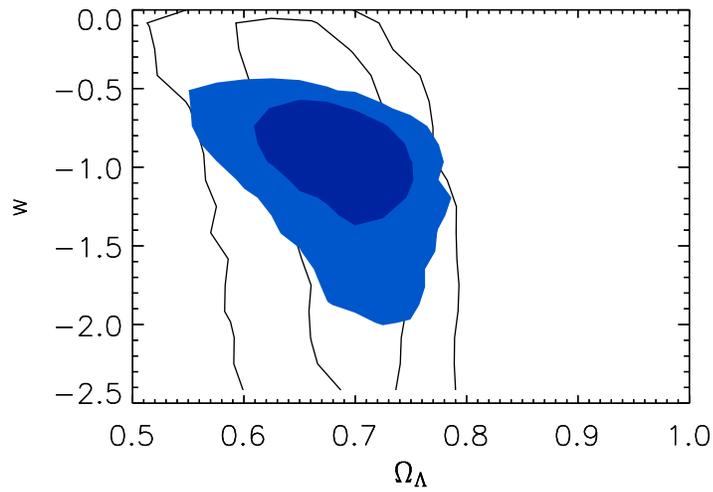}
\caption{\label{fig:wol} Constraints in the
     $w$-$\Omega_{\Lambda}$ plane for an open-w(z)CDM
     model. Solid  contours are from the WMAP5 and $H_0$
     datasets, while the solid contours use $H(z)$ information.}
\end{center}
\end{figure}

\begin{figure}
\begin{center}
\includegraphics[width=0.8\columnwidth]{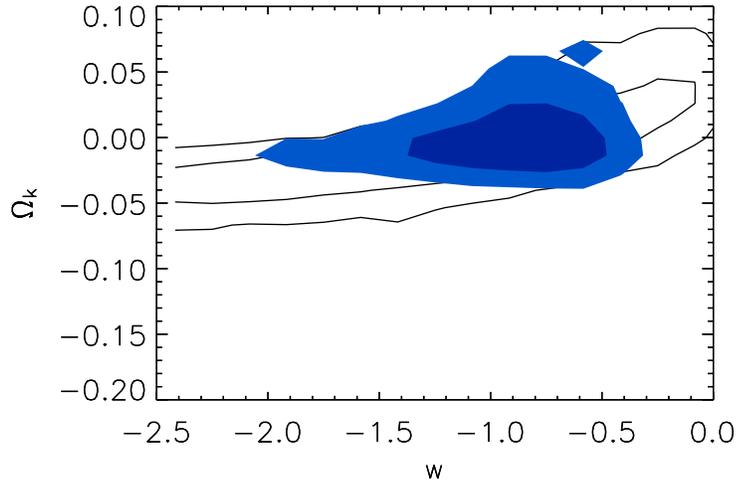}
\caption{\label{fig:wk} Constraints in the $w$-$\Omega_{k}$
     plane for an open-w(z)CDM model. Solid contours are from
     the WMAP5 and $H_0$ (HST key project; \cite{H0}) datasets,
     while the solid contours use $H(z)$ information.} 
\end{center}
\end{figure}

\begin{figure}
\begin{center}
\includegraphics[width=0.8\columnwidth]{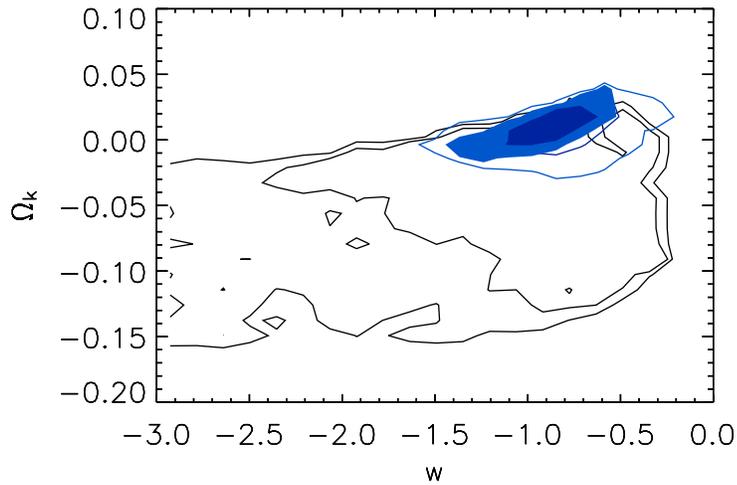}
\caption{\label{fig:w} Constraints on $w$, assuming it constant,
     for an open-wCDM model. The solid empty contours are
     obtained using WMAP5 data; the solid blue contours
     add the $H_0$ constraint from Ref.~\protect\cite{Riess09};
     the filled contours also include $H(z)$ information.}
\end{center}
\end{figure}

\begin{figure}
\begin{center}
\includegraphics[width=0.8\columnwidth]{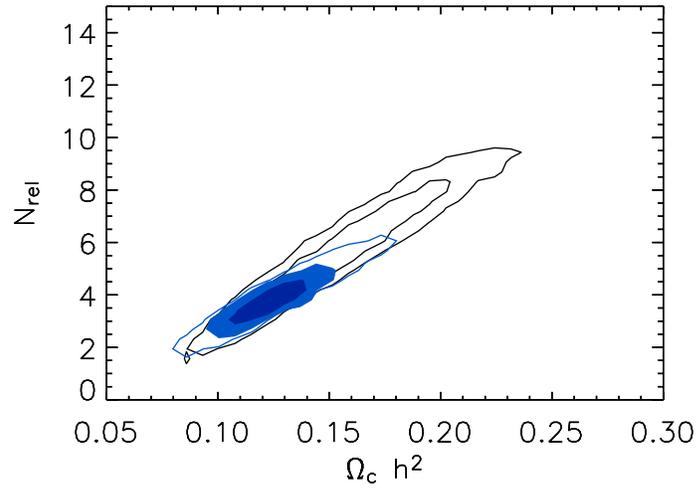}
\caption{\label{fig:neu1} Constraints on the number of
     relativistic species. Solid black contours are with WMAP5
     data only; blue transparent contours are for WMAP5 data and
     $H_0$;  and blue filled contours are WMAP5+$H_0$.}
     t
\end{center}
\end{figure}

\begin{figure}
\begin{center}
\includegraphics[width=0.8\columnwidth]{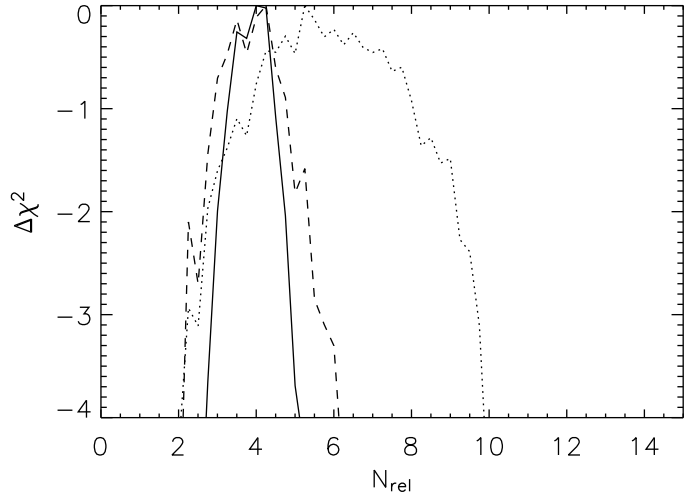}
\caption{\label{fig:neu2} Constraint on the number of
     relativistic species using WMAP5 (dotted);
     WMAP5 with the $H_0$ measurement of Ref.~\cite{Riess09}
     (dashed); and WMAP5+$H_0$+$H(z)$ data (solid line).}
\end{center}
\end{figure}

\begin{figure}
\begin{center}
\includegraphics[width=0.8\columnwidth]{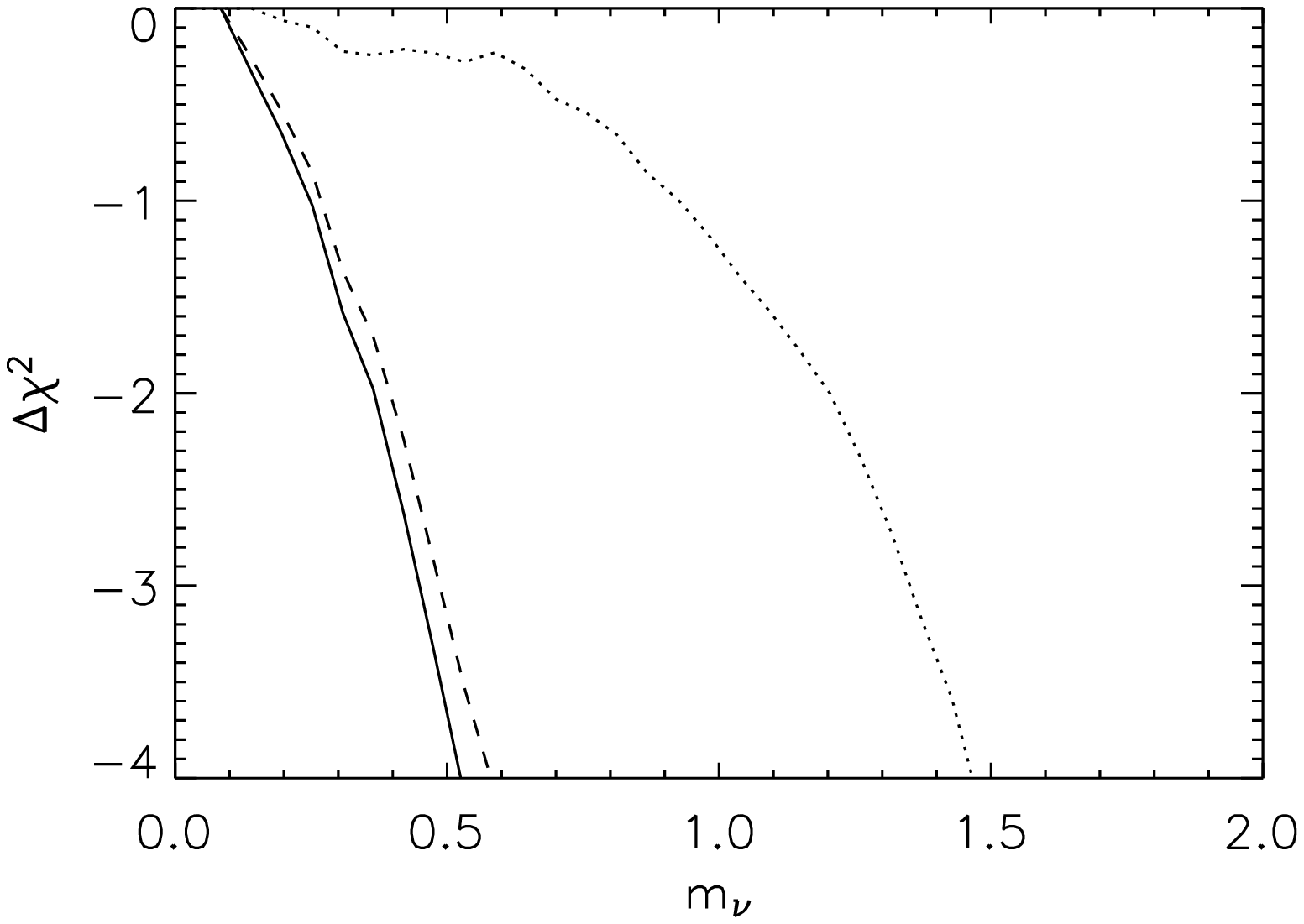}
\caption{\label{fig:neu3} Constraints on the total mass of
     relativistic species, in eV,  using the WMAP5 data only
     (dotted line); including the new $H_0$ measurement from
     Ref.~\cite{Riess09} (dashed line); and adding the $H(z)$ data
     which improves the $H_0$ constraints by $\sim 20\%$. The
     total mass of relativistic species is constraint to be $<
     0.2$~eV at $1\sigma$.}
\end{center}
\end{figure}

We have obtained Monte Carlo Markov chains for the WMAP 5-year
data \cite{WMAP5, DunkleyWMAP5}, from the LAMBDA web site
({\tt www.lambda.gsfc.gov}) and we importance-sampled them with
the $H(z)$ data. For the non-flat dark-energy models we run
independent Monte Carlo Markov chains.
Figs.~\ref{fig:wwa}--\ref{fig:neu3} illustrate how the $H(z)$
data help to constrain cosmological parameters beyond the standard
model.
 
We start by considering constraints on dark energy where the
dark-energy equation of state is parameterized as $w(z) = w +
w_a (1-a)$ and where we allow for arbitrary curvature; we call
this model the open-w(z)CDM model. Fig.~\ref{fig:wwa} shows the
improvement in the $w$,$w_a$ plane over CMB+$H_0$ constraints
where the $H_0$ constraint comes from the Hubble Key Project
\cite{H0}. Fig.~\ref{fig:wol} shows constraints in the
$w$-$\Omega_{\Lambda}$ plane, and constraints on curvature for
the above parameterization of the equation of state are shown in
Fig.~\ref{fig:wk}.  Adding the $H_0$ determination of
Ref.~\cite{Riess09} visibly improves the constraints only in
Fig.~\ref{fig:wol}, shown as the darker filled contours.
 
If we impose an equation-of-state parameter constant in
redshift, then the constraints on the curvature improve
noticeably, as seen in Fig.~\ref{fig:w}. The  wide
transparent  contours are from WMAP5 data alone (with a top hat
prior on $H_0<100~{\rm km}~{\rm s}^{-1}~{\rm Mpc}^{-1}$); the smaller transparent contours  are for
WMAP5 data and the Ref.~\cite{Riess09} $H_0$ determination;
filled contours also add $H(z)$ data. 

We next turn our attention to the constraints 
obtained on neutrino properties. Figs.~\ref{fig:neu1} and
\ref{fig:neu2} show the constraints that can be obtained on the
number of relativistic species, $N_{\rm rel}$. There is a noticeable
improvement over previous constraints. We find $N_{\rm rel}
= 4 \pm 0.5$ at the $1\sigma$ level. Finally, we present
constraints on the total mass of relativistic species. As shown
in Fig.~\ref{fig:neu3}, the best constraints we obtain using the
$H(z)$ data is $m_{\nu} < 0.2$~eV, which is an improvement of
$\sim 20\%$ over the constraints obtained using the new $H_0$
measurements by Ref.~\cite{Riess09} and a factor of two
improvement over our previous study \cite{Figueroa}.

\begin{figure}
\begin{center}
\includegraphics[width=0.8\columnwidth]{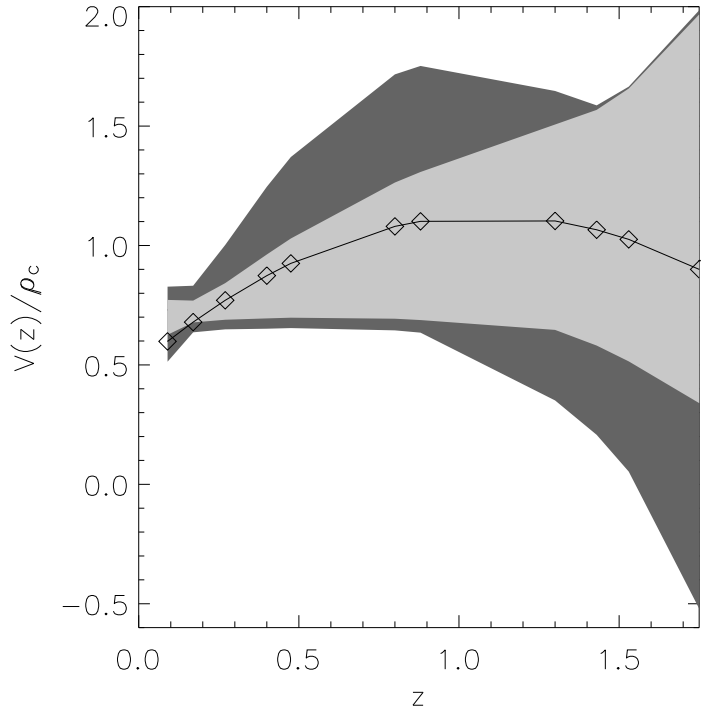}
\caption{\label{Fig:vz} Reconstructed dark-energy potential
     $V(z)$ using an expansion in Chebyshev polynomials
     truncated at second order (i.e., three coefficients). To
     produce these constraints, we have used the $H(z)$ data
     presented here, the $H_0$ determination of
     Ref.~\cite{Riess09}, and  a weak $\Omega_m$ prior ($\Omega_m=0.27
     \pm 0.07$).  Gray regions show the $1\sigma$ and $2\sigma$
     confidence regions; the solid line is the best fit $V(z)$.
     The diamonds show the redshifts at which $H(z)$ has been measured.}
\end{center}
\end{figure}

Finally, we consider dark energy that is described by a slowly
rolling scalar field and we attempt a parametric reconstruction
of its potential $V(z)$.  For this purpose we assume the
Universe to be spatially flat and we follow the steps outlined
in Ref.~\cite{svj}. We approximate a generic potential as an
expansion in Chebyshev polynomials and we truncate the expansion
at second order (i.e., we have three dark-energy parameters).
The quantity $[H(z)]^2$ is  a suitably weighted integral in
redshift of $V(z)$ (see Ref.~\cite{svj}; Eq. (29)).  We then use
the $H(z)$ data presented here, the $H_0$
determination of Ref.~\cite{Riess09}, and  a weak $\Omega_m$
prior ($\Omega_m=0.27 \pm 0.07$0. The resulting reconstructed
$V(z)$ is shown in Fig.~\ref{Fig:vz}. The reconstruction is fully
consistent with a cosmological constant.

\section{Conclusions} 
\label{sec:conclusions}

We report on a new measurement of the expansion history
obtained from the ages of passively-evolving galaxies in galaxy
clusters  at $z<1.0$.  Such
observations provide a promising new cosmological constraint,
particularly for understanding the evolution of the dark-energy
density, with a relatively modest allocation of telescope time,
effectively just two good nights at Keck.
The current measurements already provide valuable constraints,
and the success of this first campaign should motivate further
measurements along these lines as well as a more intensive
investigation of the theoretical underpinnings of the
calculations and the associated uncertainties.  Further, there
has been significant advancement in the last few years in modeling
stellar populations of LRG galaxies \cite{Wolf, Maraston08,Coelho} and
the differential technique has been recently applied
very successfully to determine the metallicity of LRGs
\cite{Walcher}. It may be that
with additional effort on both the theoretical and observational
side that the differential-age technique may ultimately provide
an important new dark-energy avenue which complements supernova
searches, weak lensing, baryon acoustic oscillations, and
cluster abundances.

\section*{Acknowledgments}

The work of DS was carried out at the Jet Propulsion Laboratory,
operated by the California Institute of Technology under a contract
with NASA.  The work of RJ and LV is supported by funds from the
Spanish Ministry for Science and Innovation AYA 2008-03531 and the
European Union (FP7 PEOPLE-2002IRG4-4-IRG\#202182).  MK was supported
by DoE DE-FG03-92-ER40701 and the Gordon and Betty Moore Foundation.
We thank the members of the SPICES and VVDS teams for making their
spectroscopic data publicly available, and we acknowledge the use
of the Legacy Archive for Microwave Background Data Analysis (LAMBDA).
Support for LAMBDA is provided by the NASA Office of Space Science.
Finally, the authors wish to recognize and acknowledge the very
significant role and reverence that the summit of Mauna Kea has
always had within the indigenous Hawaiian community. We are most
fortunate to have the opportunity to conduct observations from this
mountain.

\section*{References}
\bibliographystyle{JHEP}

\providecommand{\href}[2]{#2}\begingroup\raggedright

\end{document}